\documentclass[11pt, a4paper, copyright, goog]{google}

\usepackage[numbers, sort&compress]{natbib}
\usepackage{longtable}
\usepackage{tikz}
\usetikzlibrary{arrows.meta,positioning}
\usepackage{multirow}
\usepackage{array}
\usepackage{float}

\paperurl{}
\correspondingauthor{}
\reportnumber{}
\uselogo{}

\title{Characterizing the Quality Profile of AI-Generated C++ in Production}
\author{Michael Tran, Fred Lewis, Kun Yang, Saksham Thakur, Aditya Kini, Aditya Patil, Milad Hashemi, Parthasarathy Ranganathan}
\affil{Google, Mountain View, USA}

\begin{abstract}
The widespread integration of AI coding assistants offers undeniable boosts to engineering velocity. Yet, recent studies point to a growing trade-off, revealing persistent challenges with code quality and maintainability. Industry leaders, including frontier AI labs, echo these concerns. As large language models are increasingly relied upon to author production code, understanding their impact on shipped software quality has become a critical priority. However, assessing these effects in industrial workflows remains difficult due to challenges with observability. We study the impact of AI-generated code on production quality within a large enterprise that operates multiple global products relied upon by billions of users daily. Driven by this immense scale and user trust, the organization deeply values code quality and has built thorough observability for every line of code deployed into production, enabling us to overcome industry-wide measurement barriers to assess these effects.

This study presents a large-scale empirical analysis of AI-generated C++ code from April 2025 to April 2026, tracking 3.52 million code changes across this enterprise's massive, brownfield codebase. The core purpose is to understand the quality, performance, and maintenance characteristics of AI-generated code compared to human-written code in a production environment of this scale. We find that AI-generated C++ code has a distinct quality profile, showing a higher rate of interface and coupling burdens, copy and allocation overheads, and a stylistic reliance on explicit loops over optimized standard APIs. These issues translate into tangible downstream costs, including increased review effort and a 5-8\% increase in compute resource consumption. However, we also demonstrate that providing models with targeted, taxonomy-informed feedback can mitigate these effects, leading to an 11.1\% reduction in targeted static analysis warnings and improved computational efficiency.

\end{abstract}

\begin{document}

\maketitle

\section{Introduction}

Generative AI has moved from a laboratory capability to a routine part of software development, with code completion, text-to-code generation, and agentic editing now embedded in programming workflows. Field and enterprise studies report speedups from these tools \cite{peng2023productivity,paradis2025speed}, and case-study evidence reports similar productivity gains in everyday development \cite{coutinho2024productivity}. Diary and interaction studies also report trust, steering, and verification costs when developers use generated code in real tasks \cite{butler2025deardiary,pereira2025exploringgenai,brown2024trust,barke2023grounded,mozannar2024reading}. In our setting, AI-generated changes comprised nearly 70\% of submitted code with known authoring provenance by the close of our study window (April 1, 2025--April 1, 2026). This includes code generated through a diverse set of interaction modes, ranging from low-latency code completion models to complex multi-turn agent pipelines. Generated code can exhibit correctness, security, and maintainability weaknesses \cite{pearce2022asleep,sandoval2023lost,siddiq2022codesmells,wang2025codeerrors}. Other studies report performance, efficiency, and API-modernization weaknesses \cite{li2024performance,erhabor2025cppruntime,rajput2026correctnessefficiency,apsan2026energyefficient,wang2025deprecatedapi}. Completion rate or task speed leaves out production effects, including whether generated code survives review, creates maintenance burden, or contributes to compute cost at deployment scale \cite{sadowski2015tricorder,sadowski2018modern,ivankovic2024productive}.

% Evaluating AI-generated code in production requires a fundamentally different approach to measuring code quality. In practice, automated tools must provide targeted feedback that integrates into the developer's daily review workflow \cite{sadowski2015tricorder,sadowski2018modern}. Generated code appears as edits inside reviews, tests, static analyzers, and deployment pipelines, where it may be accepted, revised, moved, or removed before landing \cite{ivankovic2024productive,frommgen2024reviewcomments,vijayvergiya2024autocommenter}.

Evaluating AI-generated code in production requires a fundamentally different approach to measuring code quality. First, evaluation systems must associate generated code edits across a complex tooling ecosystem, spanning code reviews, tests, static analyzers, and deployment pipelines \cite{sadowski2015tricorder,sadowski2018modern}. Furthermore, these measurements must track the code across a lifetime of edits, capturing how the generated lines are accepted, heavily revised, moved, or removed before final submission \cite{ivankovic2024productive,frommgen2024reviewcomments,vijayvergiya2024autocommenter}. Because a warning may disappear during review or remain in the submitted revision and a function-level cost signal may appear only after deployment, measuring quality in this setting requires fine-grained authorship and aligned line-, change-, and function-level outcomes. In large production environments, C++ remains central in performance-sensitive systems. In these contexts, choices around memory movement, container usage, API modernity, and low-level abstractions affect both maintainability and runtime efficiency. C++ is also a useful stress case because production code combines mature static analysis with local performance-relevant idioms, such as copying versus moving, declaration placement, container construction, and API selection. Indeed, LLM-automated optimization frameworks like ECO \cite{lin2025eco} are already being deployed specifically to refactor these exact idioms at warehouse scale. Prior C++ and non-functional studies motivate the same concern because correctness alone does not characterize engineering burden in performance-sensitive systems \cite{erhabor2025cppruntime,rajput2026correctnessefficiency}. We need to know whether AI-generated C++ has a recognizable issue profile, whether that profile appears after review and deployment, and whether it is large enough to justify intervention.

Most prior work does not observe the production lifecycle because existing studies evaluate generated code in controlled prompting settings, benchmark tasks, repository snapshots, or online evaluations on real-world completions that stop before code reaches review \cite{yetistiren2022copilotquality,jamil2025quality,liu2023refining,izadi2024practicaleval,hellendoorn2019completionfails}. Non-functional studies usually analyze standalone tasks whose authorship is known by construction \cite{li2024performance,erhabor2025cppruntime,rajput2026correctnessefficiency,apsan2026energyefficient,wang2025deprecatedapi}. In production repositories, authorship is often mixed within the same change, while review, static-analysis, and compute effects are observed at line, change, and function levels, complicating measurement of where AI-generated code appears, what issues characterize it, and which categories are most relevant downstream. We observe AI generation during authoring and follow the generated code through static findings, code review, and function-level compute outcomes. Interaction- and review-centric studies provide complementary evidence about steering costs, sense-making, AI-assisted review behavior, and human-in-the-loop agent workflows, but they do not analyze mixed-authorship production code alongside downstream review and runtime outcomes \cite{barke2023grounded,mozannar2024reading,cihan2025automatedreview,takerngsaksiri2025hula}.

This gap in production-lifecycle analysis is particularly critical given the growing consensus across academia and industry regarding the trade-offs of AI-assisted development. While general surveys chart massive growth in code generation technologies \cite{huynh2025large}, recent data clearly highlights accompanying shifts in code quality and maintainability \cite{sonar2026state}. Static analyses further highlight variations in security and reliability across different platforms \cite{munoz2024comparative,al2025investigating}. In industrial software engineering, teams are increasingly navigating a balancing act: capitalizing on high initial development velocity while managing the subsequent risk of technical debt and maintenance overhead \cite{anderson2025hidden, he2025speed}. Even as frontier models continue to evolve toward parity with human developers\cite{anthropic2026when} managing the structural footprint of AI-assisted code remains a central challenge. Yet, despite this broad consensus on the quality-velocity trade-off, the structural and performance characteristics of AI-generated code have not yet been evaluated at scale within a large-scale brownfield industrial codebase.

We analyze AI-generated C++ in production, spanning change structure, static issue categories, source-level efficiency measures, and downstream outcomes in a large technology company with a monorepo-based development environment, centralized review, and post-submit operational monitoring. Using authoring-time provenance, we observe AI generation across submitted changes, final submitted versions, and deployed functions. Our analysis covers 3.52 million submitted changes and a focused C++ slice of 10.46 million lines of code with informative authoring provenance for the static-analysis study. We use these data to measure adoption at production scale, identify concentrated upstream code properties, estimate their associations with review, reliability, and compute outcomes after adjustment, and test targeted feedback on  functions.

We organize the empirical analysis around four research questions. \textbf{(RQ1)} How is AI generation distributed across submitted changes in the development workflow? \textbf{(RQ2)} What upstream code properties characterize AI-generated and human-written C++ changes, including change structure, static issue categories, and source-level efficiency measures? \textbf{(RQ3)} How do the upstream properties identified in RQ2 relate to later review, reliability, and compute outcomes? \textbf{(RQ4)} Can taxonomy-informed feedback reduce the highest-priority issue categories on  AI-generated C++ functions? This paper makes four contributions.
\begin{itemize}
\item We present a longitudinal study of AI-generated C++ in production code changes using authoring-time provenance, characterizing how AI generation is distributed across submitted changes during the study window.
\item We compare AI-generated and human-written C++ changes on upstream code properties, including change structure, a taxonomy-grounded static issue profile, and source-level efficiency measures.
\item We compare downstream outcomes between AI-generated and human-written C++ code and estimate how earlier structural, static, and efficiency-related measures are associated with those outcomes.
\item We evaluate whether taxonomy-informed feedback reduces targeted static findings on  AI-generated C++ functions.
\end{itemize}

\section{Related Work}\label{sec:related-work}

Empirical studies of LLM-generated code usually measure code before it is used in development workflows. Early Copilot and ChatGPT studies reported uneven correctness and code quality across tasks, models, and prompts \cite{yetistiren2022copilotquality,jamil2025quality}. Studies of real completion behavior show that benchmark completions do not fully capture what developers accept in practice \cite{hellendoorn2019completionfails}. Fine-grained error taxonomies further show that generated code can fail through syntactic, semantic, and repair-related patterns \cite{wang2025codeerrors}. Work on non-functional properties reports security weaknesses, runtime inefficiency, and C++ performance costs \cite{pearce2022asleep,erhabor2025cppruntime}. These studies establish that pass rates miss important quality variation, but most evaluate standalone tasks, repository snapshots, or generated solutions with authorship fixed by study design. In production repositories, generated code can be edited by developers, mixed with human-authored code, revised during review, and observed again only through submitted code or post-submit outcomes. That lifecycle changes both the measurement unit and the interpretation of quality. A finding observed on an initial submitted line may disappear before landing, while a source-level performance pattern may become visible only after deployment. Our study addresses this gap by observing AI generation during authoring and following submitted C++ changes through static findings, review, reliability, compute outcomes, and targeted feedback, where review can alter issue burden and downstream signals can appear at a later unit than the line where the generated code was written.

% Related work on developer productivity and AI-assisted workflow explains why this production setting is important.

Related work on developer productivity and AI-assisted workflow motivates the importance of analysis in a production setting. Controlled and enterprise studies report faster task completion, while interaction studies describe steering, inspection, and trust costs \cite{peng2023productivity,barke2023grounded}. Industrial work on LLM-based review and human-in-the-loop agent deployments shows that AI systems can affect review time, failure trajectories, and developer oversight \cite{cihan2025automatedreview,takerngsaksiri2025hula}. These process studies focus on task time, interaction behavior, review-tool use, repository activity, or benchmark-solving traces, so they do not characterize the source-level issue profile of submitted production code. Submitted code also raises an attribution problem because final code may contain model output that developers edited, interleaved, or partly rewrote. Detectors of AI-generated source code generalize poorly across models and settings \cite{suh2025detecting}, and memorization studies show why origin can be hard to infer from final code alone \cite{alkaswan2024memorisation}. Our setting avoids post hoc attribution by using authoring-time provenance as a measured property of the development process. The study also builds on industrial systems for static analysis, code review, and actionable feedback. Tricorder and Modern Code Review show how analysis and review operate under large-codebase workflow constraints \cite{sadowski2015tricorder,sadowski2018modern}. Productive Coverage and recent LLM-based review systems show that quality signals become more useful when they are prioritized by practical relevance and presented as developer-facing feedback \cite{ivankovic2024productive,ngassom2024verification}. Existing work motivates our evidence standard, but it does not characterize provenance-observed AI-generated C++ after industrial review, relate taxonomy-level static findings to downstream review, reliability, and compute outcomes, or test whether high-priority categories respond to targeted feedback using authoring-time provenance instead of post-submit reconstruction.

\begin{figure*}[t]
\centering
\includegraphics[width=\linewidth]{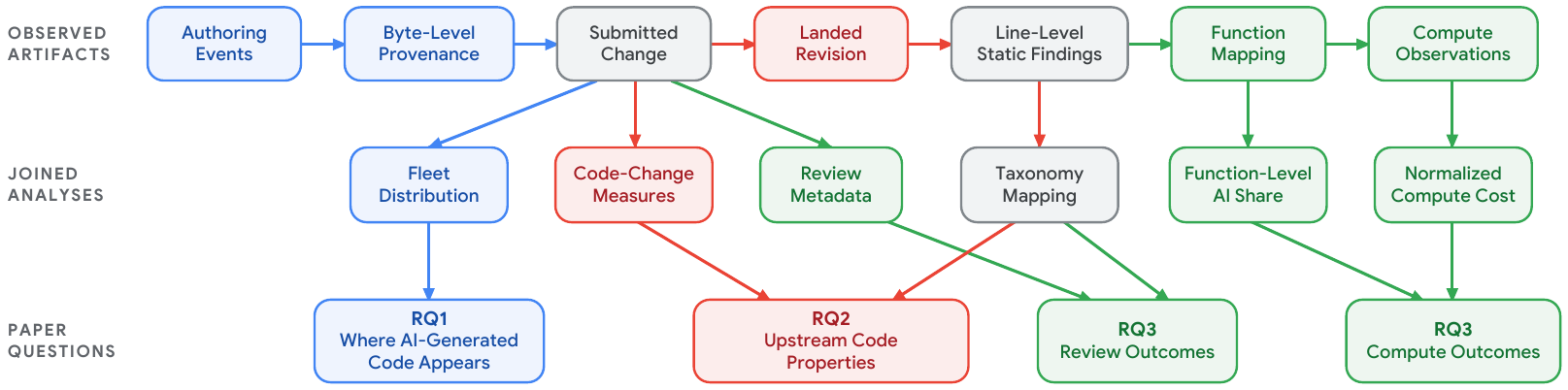}
\caption{Study overview. Provenance is observed during authoring, aggregated to submitted changes, projected to landed revisions, and joined to static findings together with review and operational outcome signals. Arrows indicate aggregation, projection, and cross-surface joins. RQ1 uses submitted changes fleet-wide, RQ2 characterizes upstream C++ code properties, and RQ3 links those upstream signals to review and compute outcomes.}
\label{fig:study-overview}
\end{figure*}

\section{Study Design}\label{sec:study-design}
\subsection{Dataset and Study Scope}

The study is conducted in a large technology company with a multi-language monorepo, centralized review, and standard build and static-analysis pipelines. The analysis follows submitted code across 5 levels of observation, from organization-wide provenance to C++ change structure, line-level static findings, review and reliability outcomes, and function-level compute observations. A  intervention then tests targeted feedback on categories selected from the upstream analyses. We exclude generated outputs, vendor code, and code areas that bypass the standard pipeline, consistent with prior industrial studies of large-scale review and static-analysis ecosystems \cite{sadowski2015tricorder,sadowski2018modern,ivankovic2024productive}. The organization-wide analysis covers submitted changes from April~1,~2025 to April~1,~2026, using all languages represented in the provenance data during that window. The final analysis includes 3.52 million submitted changes overall, and the focal C++ sample covers 10.46~million lines of code with informative authoring provenance, excludes most tests from static-finding comparisons, and contains 350k reviewable changes covered by static analysis. Function-level compute analyses require source-range-to-function mapping and normalized post-submit compute measurements from production monitoring, producing 70k observations. Comparative compute analysis observes cohorts of approximately 25k AI-heavy and 12k human-typed functions, and further stratifies by edit volume, resulting in approximately 6k matched functions in each cohort.

\begin{figure}
    \centering
    \includegraphics[width=\linewidth]{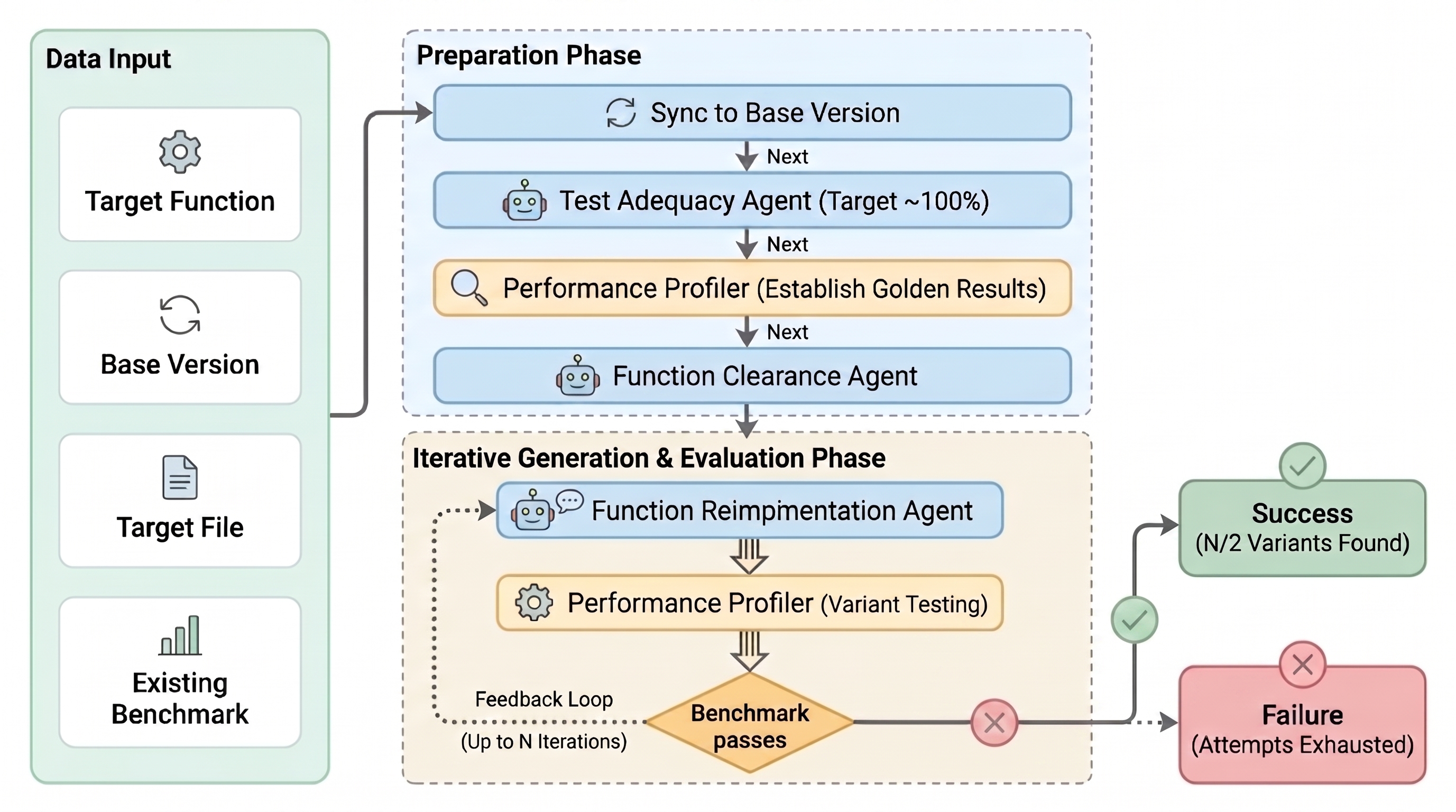}
    \caption{Evaluation dataset generation process for the intervention analysis.}
    \label{fig:rq4_gen}
\end{figure}

Submitted changes are the unit for organization-wide distribution, upstream code-property, review, and reliability analyses, while functions are the unit for compute analyses. We aggregate byte-level provenance within each change to obtain AI-generation share across months, languages, interaction modes, and anonymized organizational slices. The intervention uses a synthesized benchmark of 50 C++ functions with at least one target-category finding and micro-benchmarks for validating computational behavior. We remove each original function body while preserving inputs, outputs, and intended logic, then ask the coding agent to reimplement the function under the prompt conditions in Section~\ref{sec:intervention}, as shown in Figure~\ref{fig:rq4_gen}.

\subsection{Provenance, Revision Mapping, and Analysis Units}

\noindent\textbf{Provenance and projection.} We observe authorship provenance on individual changed bytes during authoring, and then project these byte-level annotations onto the lines, functions, and static findings of the final submitted snapshot used by later analyses. We aggregate consistently logged tool features into AI-generation and human-authoring categories, pooling inline completion, conversational generation, agentic editing, and transformation-based editing into 1 AI-generation signal.
% For each informative byte $b$, bytes with only AI-generation features contribute $(1,0)$, bytes with only human-authoring features contribute $(0,1)$, and bytes carrying both are handled by the overlap policy, with $(0.5,0.5)$ used under the split-weight policy. Bytes with neither feature family are excluded from the denominator. Robustness checks repeat the main comparisons with overlapping bytes excluded.

% \noindent\textbf{Cohorts and analysis units.} For a line $\ell$, $AIShare(\ell)$ is the weighted fraction of informative bytes attributed to AI-generation features. A static finding inherits $AIShare$ from the matched line or line span, using a length-weighted mean when the finding spans multiple lines. Function-level compute analyses aggregate AI and human informative byte counts over the enclosing function and project static findings to the same function to derive category covariates. For descriptive figures, `AI-generated' units have share $>0.5$, `human-written' units have human share $>0.5$ and are not AI-generated, and `mixed AI/human' units cover all remaining cases, while inferential models use continuous AI-generation share.

\noindent\textbf{Cohorts and analysis units.} For a line $\ell$, $AIShare(\ell)$ is the fraction of informative bytes attributed to AI-generation features. Because static analysis tooling often attaches a finding to a narrow subset of the symptomatic code, a static finding inherits $AIShare$ from the full matched line or line span, using a length-weighted mean when the finding spans multiple lines. Function-level compute analyses aggregate AI and human informative byte counts over the enclosing function and project static findings to the same function to derive category covariates.

\subsection{Static Taxonomy}\label{sec:taxonomy-definition}

The taxonomy classifies line-level static findings in C++ final submitted snapshots. Each mapped finding is anchored to a line or line range and propagated to the enclosing function, file, and code change. Review comments, review latency, revert behavior, sanitizer outputs, and post-submit compute cost are downstream outcomes and are not part of the taxonomy. The 3-level taxonomy groups findings into quality attributes, issue categories, and issue types.

We construct the taxonomy by collapsing raw checks and aliases across tools, then assigning each finding to a primary issue category based on the developer-facing problem it signals. The initial codebook is seeded from high-frequency checks and refined through stratified samples, following taxonomy-development guidance and empirical taxonomy practice for generated-code failures \cite{usman2017taxonomies,izadi2024practicaleval,wang2025codeerrors}. 2 annotators independently code a validation sample spanning frequent checks, high-AI-share checks, and every quality attribute. Before downstream use, domain experts review the final raw-check-to-category mapping and confirm that high-volume checks retain their developer-facing meaning. The validation records raw-check coverage, validation sample size, inter-rater agreement, and unmapped-bucket size.

\begin{table}[t]
\caption{Top-level definitions for the static taxonomy used in the C++ analyses.}
\label{tab:quality-taxonomy}
\centering
\footnotesize
\begin{tabular}{@{}>{\raggedright\arraybackslash}p{0.3\linewidth} >{\raggedright\arraybackslash}p{0.6\linewidth}@{}}
\toprule
\textbf{Quality Attribute} & \textbf{Definition} \\
\midrule
\textbf{\textit{Efficiency and Resource Use}} & Avoidable runtime or resource overhead in otherwise valid code, including wasted work, needless copying, inefficient access patterns, or expensive low-level choices. \\
\addlinespace
\textbf{\textit{Correctness and Safety}} & Semantic hazards, invalid state, unsafe API use, or undefined-behavior risk. \\
\addlinespace
\textbf{\textit{Maintainability and Readability}} & Issues that increase inspection, review, or maintenance effort by making code harder to read, reason about, or evolve. \\
\addlinespace
\textbf{\textit{Modernity and API Evolution}} & Outdated language usage, obsolete APIs, or missed adoption of safer and clearer modern C++ idioms. \\
\addlinespace
\textbf{\textit{Policy, Portability, and Environment Fit}} & Poor fit with platform, policy, or deployment constraints that can already be detected statically. \\
\bottomrule
\end{tabular}
\vspace{-10pt}
\end{table}

\subsection{Metrics and Models} \label{subsec:metrics}

% \noindent\textbf{Upstream code measures.} We use added LOC, deleted LOC, add/delete ratio, files touched, analyzed C++ LOC, and new-code status as change-level descriptors and controls. The static issue profile is measured with weighted findings per analyzed KLOC, category composition, and AI/human rate ratios. Mixed-authorship findings contribute to AI and human rates in proportion to observed provenance shares. To summarize concentration, we report the absolute rate gap and the share of the positive gap represented by the top-2 categories. Issue-type ratios are reported only in follow-up analyses, and issue types with combined support below 1,000 are suppressed. Main-paper tables report ratios, shares, percentages, and support thresholds, omitting raw internal volumes.

\noindent\textbf{Upstream code measures.} We use added LOC, deleted LOC, add/delete ratio, files touched, analyzed C++ LOC, and new-code status as change-level descriptors and controls. The static issue profile is measured with weighted findings per analyzed KLOC, category composition, and AI/human rate ratios. Mixed-authorship findings contribute to AI and human rates in proportion to observed provenance shares. Issue-type ratios are reported only in follow-up analyses, and issue types with combined support below 1,000 are not included.

\noindent\textbf{Source-level efficiency measures.} We measure loop constructs, standard-library/API usage, move-related warnings, container-insertion warnings, map-access warnings, repeated-work warnings, and low-level implementation-overhead warnings on the same C++ submitted-code sample. Each measure is summarized by observed signal, AI/human contrast, denominator, and its relationship to compute modeling. We include a measure in the compute models only when it meets the predefined support threshold, remains directionally stable under normalized denominators, and has a concrete relationship to runtime cost.

% \noindent\textbf{Review, reliability, and compute outcomes.} Review and reliability outcomes are measured at the code-change level using categories computed on the initial submitted snapshot. The primary review outcomes are total comments, blocking comments, and time to merge, with reviewer iterations and submit attempts reported as secondary process outcomes and build failure, sanitizer findings, and revert rate reported as reliability outcomes. These review measures follow industrial studies that use comments, latency, and blocking interactions as workflow outcomes \cite{sadowski2018modern,frommgen2024reviewcomments,vijayvergiya2024autocommenter}. Rather than relying on parametric regression models, we evaluate these outcomes through stratified cohort comparisons. We report effect sizes as relative rates or percent changes, accounting for confounding factors by stratifying across month, change size, coarse organizational slice, anonymized author or team controls, and new-code status where available. Function-level compute comparisons use weighted AI-generation share, projected categories, and source-level efficiency measures. The main compute outcome is normalized compute cost, and function execution frequency is used as both a control and a stratification variable. To isolate the impact of AI-generated code, our primary specification relies on within-service matched comparisons, pairing AI-heavy and human-written functions with similar execution profiles and product contexts.

\noindent\textbf{Review, reliability, and compute outcomes.} Review and reliability outcomes are measured at the code-change level using categories computed on the initial submitted snapshot. The primary review outcomes are total comments, blocking comments, and time to merge, with reviewer iterations and submit attempts reported as secondary process outcomes and build failure, sanitizer findings, and revert rate reported as reliability outcomes. These review measures follow industrial studies that use comments, latency, and blocking interactions as workflow outcomes \cite{sadowski2018modern,frommgen2024reviewcomments,vijayvergiya2024autocommenter}. Rather than relying on parametric regression models, we evaluate these outcomes through stratified cohort comparisons. We report effect sizes as relative rates or percent changes, accounting for confounding factors by stratifying across month, change size, coarse organizational slice, anonymized author or team controls, and new-code status where available. Function-level compute comparisons utilize longitudinal resource tracking across two distinct function cohorts classified by AI-generation share, measuring utilization growth indexed to a baseline at the start of our study window. Our primary specification stratifies these cohorts by their respective edit volumes. To isolate relative resource consumption, the final outcomes are normalized against the growth of total application compute, denominating CPU cost and heap-resident memory strictly as a percent-of-application share.

\noindent\textbf{Execution profile shifts.} To quantify how upstream structural code properties alter runtime behavior over time, we conducted a longitudinal CPU shift analysis comparing production profiling data from the first and last months of the study window. Using the established cohorts of AI-heavy and human-written functions, we partitioned total compute cycles, spanning the focal functions and their direct callees, into distinct categories. Execution time was classified into imperative signals, which include direct on-CPU processing, object constructors, vector operations, and map lookups; and declarative signals, capturing cycles spent calling shared, optimized library implementations (e.g., \verb|absl| and \verb|gtl|). By calculating the change in the percentage of total CPU cycles allocated to these categories, we determined the net shift in the runtime execution profile of each cohort.

\noindent\textbf{Intervention outcomes.} The intervention evaluates taxonomy-informed feedback on the 50-function benchmark using 3 prompt stages and 3 independent runs per stage, producing 450 generated implementations before validation exclusions and following prior work on LLM-based review support \cite{ngassom2024verification,takerngsaksiri2025hula}. The primary outcome is targeted static finding count, and the secondary outcome is $R_{\mathrm{eff}}$, a benchmark-based score computed from CPU instruction count and memory usage relative to the original implementation. The score maps the weighted instruction/memory comparison to 3 values, with 1 indicating improvement, 0.5 parity, and 0 regression. Instruction count receives 2$\times$ the memory weight because execution latency is the main cost channel in the benchmark.

\section{Results}\label{sec:results}
\subsection{\textbf{RQ1:} Distribution and Trends}\label{sec:fleet-distribution}

\begin{figure*}[t]
\centering
\includegraphics[width=0.7\textwidth, keepaspectratio]{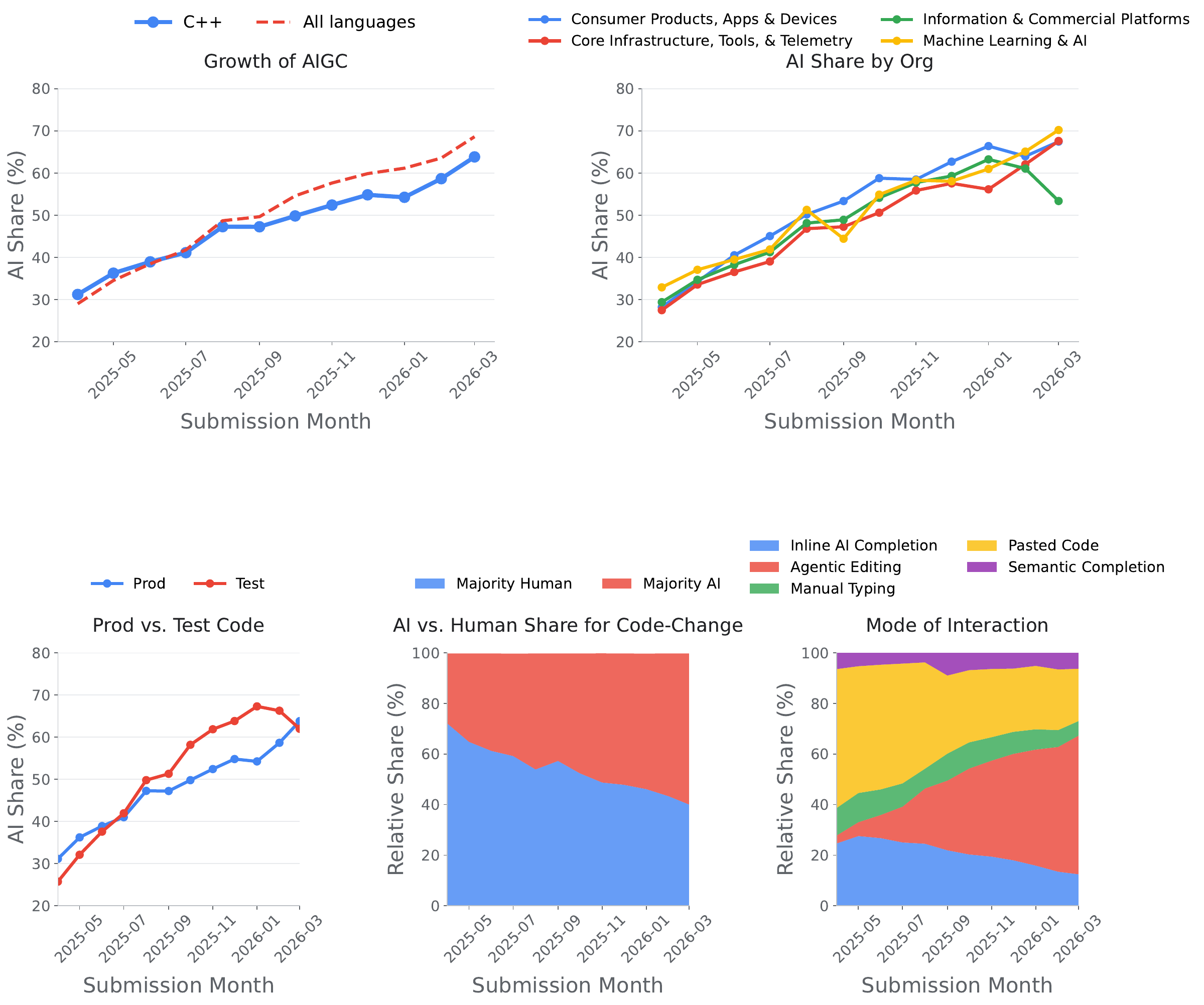}
\caption{Distribution of AI-generated code during the study window.}
\label{fig:fleet-distribution}
\vspace{-10pt}
\end{figure*}

Figure~\ref{fig:fleet-distribution} shows the organization-wide distribution of AI-generated code during the study window. Provenance instrumentation becomes available from March~2025 onward, but the reported trend and cohort summaries in this section use only submitted changes from April~1,~2025 to April~1,~2026. Each submitted change in that window inherits an AI-generation share aggregated from its annotated bytes, and the figure summarizes those submitted-change shares across languages, months, anonymized organizational slices, and interaction modes. AI-generated code represents a rapidly growing share of submitted code with informative authoring provenance during this period, with its share across major languages rising from 28.99\% in April~2025 to 68.62\% in March~2026. At the submitted-change level for C++, the share of majority AI-generated changes grew from 27.65\% to 59.69\% over the same period, while majority human-written changes fell from 72.10\% to 40.04\%. These descriptive results establish where AI-generated code appears and why the C++ sample is large enough and important enough to analyze in depth.

% \begin{figure*}[t]
% \centering
% \includegraphics[width=0.8 \textwidth, keepaspectratio]{Figures/aigc.pdf}
% \caption{Distribution of AI-generated code during the study window.}
% \label{fig:fleet-distribution}
% \vspace{-0.5em}
% \end{figure*}
% \vspace{-10pt}

% \begin{figure}[t]
% \centering
% \includegraphics[width=\columnwidth]{Figures/aigc.pdf}
% \caption{Distribution of AI-generated code during the study window.}
% \label{fig:fleet-distribution}
% \vspace{-0.5em}
% \end{figure}

We also observe that the distribution changes over time, but not at the same rate in every organizational slice. Some coarse organizational slices show steady growth, while others plateau after initial adoption. For example, the overall C++ monthly share rises from 28.56\% at the start of the window to 62.80\% at the end. By the end of the window, adoption spreads steadily across the long tail of product domains, though absolute penetration varies. AI-generated code spreads beyond the earliest high-use slices, with notable concentration visible in domains like Machine Learning \& AI (reaching 70.19\%) and Consumer Products, Apps \& Devices (reaching 67.45\%). Later analyses use distributional summaries and coarse slice- and domain-level controls instead of treating the organization as homogeneous.

The interaction-mode mix also provides important context for the later results. During the study window, bytes attributed to AI generation are distributed across two primary interaction modes: Inline tooling (which includes inline completions, smart pasting, and next-edit predictions) and agentic workflows (which encompass conversational generation, transformation-based editing, automated test generation, and AI-driven refactoring). These shares are descriptive and clarify what the pooled AI-generation signal represents in this deployment setting. The pooled AI-generation signal analyzed in this study is fundamentally heterogeneous. Over the 1-year study window, the production environment deployed a shifting mix of multiple model generations, varying significantly in parameter scale, training objectives, and context-window capabilities. This includes early-window legacy architectures optimized strictly for inline latency alongside later-window conversational and agentic models. Accordingly, the results characterize the operational reality of a rolling production ecosystem rather than the standalone capability ceiling of any singular, modern state-of-the-art model.

\noindent\textbf{RQ1 summary.} We find that AI generation is already a large source of submitted code in this workflow, but its distribution is uneven across languages, organizational slices, and interaction modes, and developers substantially filter generated text before it reaches submitted-code analysis.

\subsection{\textbf{RQ2:} Upstream Code Properties}\label{sec:taxonomy}

% \begin{table}[t]
% \centering
% \caption{Contrasts between AI-generated and human-written C++ changes. Entries report median [IQR] and Cliff's $\delta$.}
% \label{tab:structural_metrics}
% \resizebox{\columnwidth}{!}{% 
% \begin{tabular}{c l l l c}
% \toprule
% \textbf{Level} & \textbf{Metric} & \textbf{AI (Med [IQR])} & \textbf{Human (Med [IQR])} & \textbf{$\delta$} \\
% \midrule
% \multirow{4}{*}{\rotatebox[origin=c]{90}{Code-change}} 
% & Lines Changed & 89 [27, 230] & 33 [8, 122] & 0.284 \\
% & Files Touched & 3 [2, 5] & 2 [1, 3] & 0.233 \\
% & New Code Ratio & 0.83 [0.51, 0.97] & 0.60 [0.14, 0.93] & 0.232 \\
% % & Test Ratio & 0.25 [0.00, 0.44] & 0.00 [0.00, 0.50] & 0.112 \\
% & Desc. Length & 522 [375, 742] & 491 [363, 699] & 0.056 \\
% \midrule
% \multirow{3}{*}{\rotatebox[origin=c]{90}{Function}} & LOC & 11 [5, 26] & 15 [7, 31] & -0.118 \\
% % & Edit Fragments & 1 [1, 3] & 1 [1, 1] & 0.344 \\
% & Complexity & 3 [1, 6] & 3 [1, 6] & -0.029 \\
% \bottomrule

% \end{tabular}%
% }
% \end{table}

\begin{table}[t]
\centering
\caption{Contrasts between AI-generated and human-written C++ changes. Entries report median [IQR] and Cliff's $\delta$.}
\label{tab:structural_metrics}
\small
\setlength{\tabcolsep}{6pt}
\renewcommand{\arraystretch}{1.05}
\begin{tabular}{@{}l l l c@{}}
\toprule
\textbf{Metric} & \textbf{AI (Med [IQR])} & \textbf{Human (Med [IQR])} & \textbf{$\delta$} \\
\midrule
\multicolumn{4}{l}{\textit{\textbf{Code-change Level}}} \\
\midrule
Lines Changed & 89 [27, 230] & 33 [8, 122] & 0.284 \\
Files Touched & 3 [2, 5] & 2 [1, 3] & 0.233 \\
New Code Ratio & 0.83 [0.51, 0.97] & 0.60 [0.14, 0.93] & 0.232 \\
Desc. Length & 522 [375, 742] & 491 [363, 699] & 0.056 \\
\midrule
\multicolumn{4}{l}{\textit{\textbf{Function Level}}} \\
\midrule
LOC & 11 [5, 26] & 15 [7, 31] & -0.118 \\
Complexity & 3 [1, 6] & 3 [1, 6] & -0.029 \\
\bottomrule
\end{tabular}
\renewcommand{\arraystretch}{1.0}
\vspace{-10pt}
\end{table}

RQ2 uses three upstream views of the same submitted-code population, covering change structure, static issue categories, and source-level efficiency measures before review, reliability, or compute outcomes are observed. Table~\ref{tab:structural_metrics} compares AI-generated and human-written C++ code changes on code-change and function-level structure, and Table~\ref{tab:category-profile} reports the category-level static issue profile on the same submitted-code sample. AI-generated C++ changes are larger and newer, but the static issue-rate gap is concentrated in a small number of categories instead of spread evenly across the taxonomy.

Table~\ref{tab:structural_metrics} shows why change structure needs to be modeled before interpreting issue rates. Relative to human-written changes, AI-generated changes touch more lines and more files, and have a higher new-code ratio, while description length changes much less. These measures keep size, novelty, and surface area from being mistaken for issue-category effects, allowing the later comparison to focus on which static categories and local implementation patterns characterize AI-generated C++.

At the quality-attribute level, Efficiency, Resource Use and Maintainability and Readability have above-parity AI/human rate ratios of 1.23 and 1.08, respectively. Modernity and API Evolution is close to parity at 1.04, while Correctness and Safety and Policy, Portability, and Environment Fit are lower in AI-generated code, with ratios of 0.94 and 0.92. The attribute-level view provides the context for the category results in Table~\ref{tab:category-profile}. Even at this coarser level, the profile is not uniform: some quality attributes carry most of the excess burden, while others contribute little or move in the opposite direction.

\begin{table}[t]
\caption{Static issue profile of AI-generated and human-written C++, grouped by quality attribute. Columns report AI/Human rate ratio and within-attribute share of weighted findings.}
\label{tab:category-profile}
\centering
\footnotesize
\setlength{\tabcolsep}{6pt}
\renewcommand{\arraystretch}{1.03}
\begin{tabular}{@{}l c r@{}}
\toprule
\textbf{Issue Category} & \textbf{Rate Ratio (AI/Human)} & \textbf{Share (\%)} \\
\midrule
\textbf{\textit{Efficiency and Resource Use}} & \textbf{1.23} & \textbf{100.0\%} \\
\quad Copy and Allocation Overhead & \underline{1.39} & 83.4\% \\
\quad Data Structure and Access Inefficiency & 0.80 & 8.9\% \\
\quad Low-Level Implementation Overhead & 0.56 & 6.7\% \\
\quad Redundant Work & 0.79 & 0.6\% \\
\quad I/O and Formatting Inefficiency & \underline{3.16} & 0.4\% \\
\addlinespace
\textbf{\textit{Modernity and API Evolution}} & \textbf{1.04} & \textbf{100.0\%} \\
\quad Legacy Language Idioms & 0.82 & 43.1\% \\
\quad Deprecated or Obsolete API Usage & \underline{1.41} & 41.2\% \\
\quad Missed Type-Safety or Ownership Evolution & 0.96 & 15.7\% \\
\addlinespace
\textbf{\textit{Maintainability and Readability}} & \textbf{1.08} & \textbf{100.0\%} \\
\quad Interface and Coupling Burden & \underline{1.15} & 75.9\% \\
\quad Dead, Redundant, or Unused Code & 0.97 & 13.4\% \\
\quad Clarity and Explicitness & 0.86 & 5.9\% \\
\quad Naming and Local Documentation Issues & 0.80 & 4.4\% \\
\quad Control-Flow and Structural Complexity & 0.92 & 0.4\% \\
\addlinespace
\textbf{\textit{Correctness and Safety}} & \textbf{0.94} & \textbf{100.0\%} \\
\quad API Misuse and Invalid Calls & 0.93 & 91.3\% \\
\quad Numeric and Conversion Hazards & \underline{1.25} & 3.9\% \\
\quad Initialization and State Validity & 0.86 & 2.4\% \\
\quad Lifetime and Ownership Hazards & 0.71 & 1.7\% \\
\quad Concurrency and Synchronization Hazards & \underline{1.01} & 0.7\% \\
\addlinespace
\textbf{\textit{Policy, Portability, and Environment Fit}} & \textbf{0.92} & \textbf{100.0\%} \\
\quad Policy or Compliance Violations & 0.91 & 87.2\% \\
\quad Portability Risks & \underline{1.01} & 12.6\% \\
\quad Operational Environment Mismatch & 0.33 & 0.2\% \\
\bottomrule
\end{tabular}
\renewcommand{\arraystretch}{1.0}
\vspace{-10pt}
\end{table}

\noindent\textbf{Concentration.} Among categories with support of at least 1,000, the largest positive absolute rate gaps come from Interface and Coupling Burden and Copy and Allocation Overhead. These two categories represent 82.21\% of the total positive absolute rate gap. We find a narrow practical target in interface/coupling checks and copy/allocation checks, while many mapped categories stay near parity or move lower in AI-generated code. Later analyses track category rates beyond an overall finding rate because the positive gap is concentrated in these developer-facing mechanisms.

Composition shows which issue classes reviewers encounter most often. Interface and Coupling Burden represents 43.96\% of the weighted findings in AI-generated code versus 40.36\% in human-written code, and API Misuse and Invalid Calls represents 29.49\% versus 30.05\%. The same two categories make up 73.45\% of weighted findings in AI-generated code and 70.41\% in human-written code. We report rate contrast and composition because a category can be overrepresented in AI-generated code while forming a small share of the total burden, and a common category can dominate weighted findings in both cohorts without being the most disproportionate.

Support changes how we read these ratios. A high ratio with sparse support marks a narrow contrast, while a high ratio paired with a large absolute rate gap is a stronger candidate for interpretation because it is both disproportionate and substantial in the total gap. Within highlighted categories, issue types under clang-tidy\cite{lattner2004llvm} definition such as \texttt{misc-include-cleaner} and \texttt{misc-definitions-in-headers} make up 98.95\% of Interface and Coupling Burden, while \texttt{runtime-missing-move} makes up 51.60\% of Copy and Allocation Overhead. These issue types translate the category profile into concrete mechanisms that reviewers and tool builders can target.

\begin{table}[b]
\caption{Upstream source-level efficiency measures in the C++ submitted-code sample. Ratios are normalized AI/human rates.}
\label{tab:upstream-compute-candidates}
\centering
\footnotesize
\setlength{\tabcolsep}{4pt}
\renewcommand{\arraystretch}{1.08}
\begin{tabular}{@{} p{0.33\columnwidth} p{0.43\columnwidth} >{\centering\arraybackslash}p{0.14\columnwidth} @{}}
\toprule
\textbf{Measure} & \textbf{Observed signal} & \textbf{Ratio} \\
\midrule
Loop usage& AST loop counts& $\sim$2.0$\times$ \\
 Standard library use& \texttt{std::*} usages&$\sim$0.4$\times$\\
Move/copy behavior & \texttt{missing-move} warnings& 1.39$\times$ \\
Container insertion & \texttt{use-emplace} warnings & $\sim$2.0$\times$ \\
Map access & \texttt{inefficient-map} warnings & $\sim$2.0$\times$ \\
Repeated work & findings per analyzed KLOC & $<$1.0$\times$ \\
\bottomrule
\end{tabular}
\renewcommand{\arraystretch}{1.0}
\vspace{-0.5em}
\end{table}

\noindent\textbf{Source-level efficiency measures.} RQ2 extends the static taxonomy with source-level efficiency measures observed before downstream outcomes. Table~\ref{tab:upstream-compute-candidates} reports what was measured and whether the normalized AI/human rate ratio is above or below parity. These quantities are upstream code properties measured on submitted or final code snapshots and do not use computational cost, heap, or any other production outcome. Measures that fail the support or normalization-stability threshold remain descriptive. Measures with sufficient support and a plausible runtime-cost interpretation are used in Section~\ref{sec:study-design} as candidate explanatory variables. We find that AI-generated code is nearly twice as likely to author loops, uses standard-library/API calls 30--40\% less often, has almost twice as many map-access warnings, and shows a higher rate of \texttt{push\_back}-related warnings. The pattern points to more explicit local control flow and fewer calls into shared library or API implementations. The downstream analysis uses \texttt{AST (loops/libs)} for loop and standard-library/API usage, \texttt{missing-move} for move-semantics-related warnings, \texttt{use-emplace} for container-insertion warnings, and \texttt{inefficient-map} for map-access warnings.

% \begin{figure}[t]
% \centering
% \includegraphics[width=\columnwidth]{Figures/rq2_gcu.jpg}
% \caption{The computing resource shift before and after the investigated period.}
% \label{fig:gcu-outcomes}
% \vspace{-10pt}
% \end{figure}

% While the source-level efficiency measures provide a static snapshot of AI behavior, these localized implementation choices directly foreshadow specific computational bottlenecks. By pairing our static findings with preliminary runtime telemetry sampled during the same observation window, we can confirm that these upstream code properties translate directly into execution overhead, as illustrated in Figure~\ref{fig:gcu-outcomes}.

% When comparing computing usage shifts between AI-generated and human-authored code, the top-shifting categories closely align with our static taxonomy. For instance, the AI's heavy reliance on \texttt{use-emplace} and \texttt{inefficient-map} warnings corresponds to substantial runtime increases in \textit{Variable constructor} (+1.255\%) and \textit{Map lookup} (+0.141\%) overheads. In contrast, human developers actively refactored their logic during this period, resulting in a significant reduction of \textit{In function local on cpu} (-2.36\%) usage, opting instead to slightly increase \textit{Shared Library function calling} (+0.230\%). Ultimately, this telemetry confirms that the verbose, localized control flow identified statically in AI-generated code directly materializes as measurable memory allocation and CPU execution overhead in production.

\noindent\textbf{RQ2 summary.} AI-generated C++ changes are larger and more new-code-heavy, and their excess static burden is concentrated in two developer-facing categories. The source-level efficiency measures add a resource-use pattern in which AI-generated code relies more on explicit local work and less on standard-library/API delegation.

\subsection{RQ3: Review, Reliability, and Compute Outcomes}
To understand the downstream impact of AI-generated code, we tracked changes through the review process and into production, comparing reliability and efficiency metrics across our authorship cohorts.

\begin{figure}[t]
\centering
\includegraphics[width=\columnwidth]{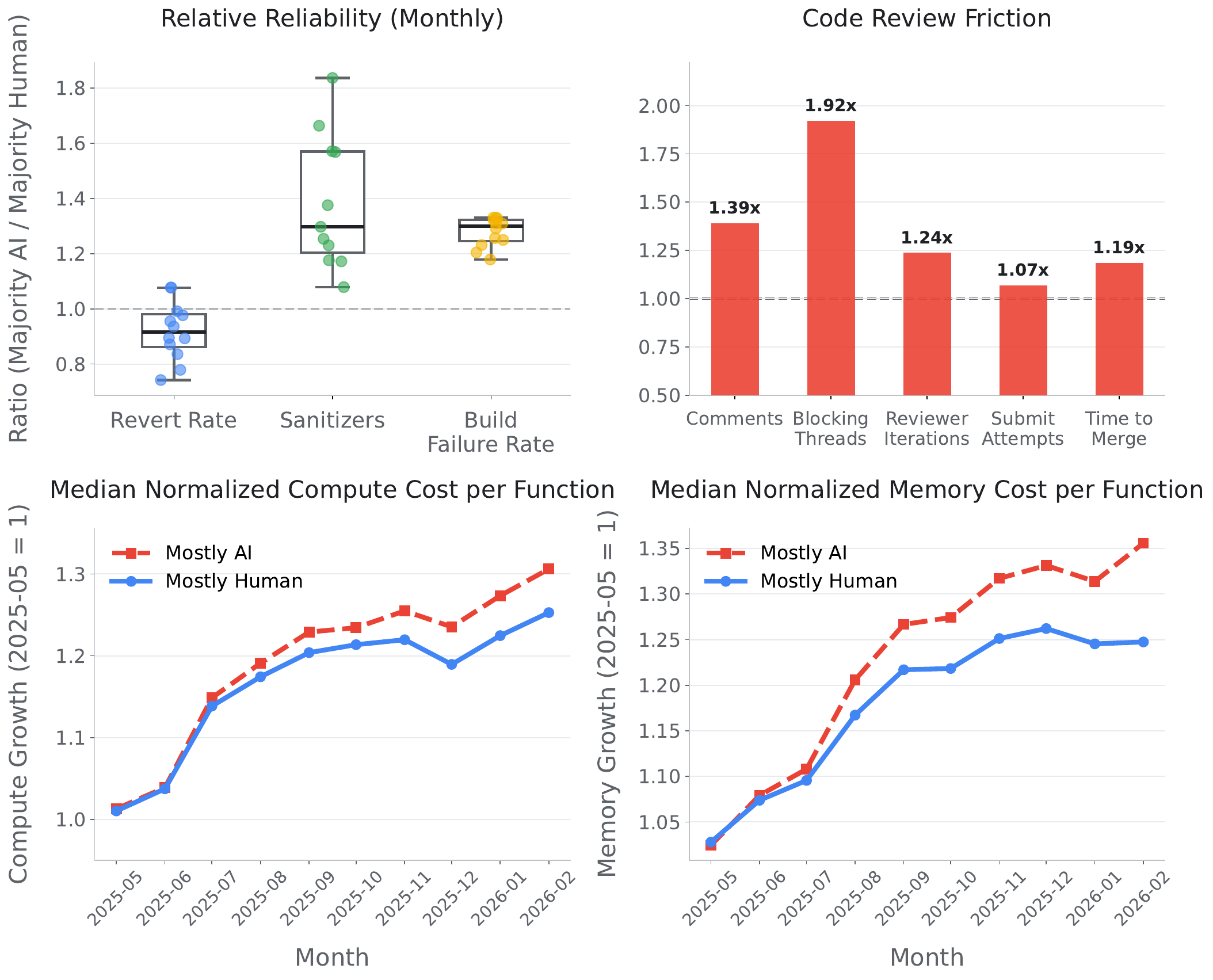}
\caption{Review, reliability, and compute outcomes for AI-generated and human-written C++ code. Panels show the descriptive downstream contrasts used to motivate the staged models: monthly ratios for revert rate, sanitizer findings, and build failure, review-friction ratios, and monthly normalized compute cost.}
\label{fig:category-outcomes}
\vspace{-10pt}
\end{figure}

\noindent\textbf{Reliability: Build Failures and Revert Rates.}
Our longitudinal cohort analysis reveals a distinct split between pre-deployment friction and post-deployment stability for AI-generated code. When examining build failure rates over the 12-month study window, the AI-generated cohort exhibited consistently higher initial instability. As shown in Figure~\ref{fig:category-outcomes}, the ratio of AI-generated to human-written build failures remained above parity (with a median ratio of roughly 1.3$\times$). 

However, AI-generated code demonstrated a consistently \textit{lower} revert rate than human-written code. Across the study period, the relative ratio for revert rates remained largely below parity (median $\sim$0.9$\times$). This indicates that while AI-generated code requires more automated pipeline cycles to stabilize, once it passes initial review and automated checks, it is less likely to require an immediate rollback than human-authored code.

\noindent\textbf{Compute and Memory Outcomes.}
When tracking code into production, we observe a growing divergence in resource consumption over time (Figure~\ref{fig:category-outcomes}, bottom panels). Over the study window, the median normalized compute cost for AI-heavy functions grew at a steeper trajectory than for mostly human-written functions. By early 2026, AI-generated compute growth reached approximately 1.31$\times$ baseline, compared to roughly 1.25$\times$ for human code (an approximate 5\% relative increase). This gap is even more pronounced in memory consumption, where AI-heavy functions reached nearly 1.36$\times$ baseline growth versus 1.25$\times$ for human functions (an approximate 8\% relative increase). This confirms that the upstream efficiency issues identified in RQ2 translate into measurable compute and memory overheads at scale.

 Table~\ref{tab:gcu-shift} reports the longitudinal CPU shift from the beginning to end of the study window, demonstrating that the source-level efficiency properties identified in Table~\ref{tab:upstream-compute-candidates} translate into measurable execution overheads in production. While Table~\ref{tab:upstream-compute-candidates} established that AI-generated code statically biases toward explicit local loops ($\sim$2.0×) and away from standard API calls ($\sim$0.4×), Table~\ref{tab:gcu-shift} reveals the dynamic consequences: AI-heavy functions exhibit a 3.46\% relative shift toward direct on-CPU execution and a 1.08\% relative decrease in delegation to declarative libraries. Furthermore, the specific static efficiency antipatterns, such as elevated warnings for missing move semantics and suboptimal container usage, manifest in localized runtime cost driving relative compute footprint increases in constructors (+1.36\%), vectors (+0.10\%), and map lookups (+0.07\%).

\begin{table}[t]
\centering
\caption{1-Year CPU Shift Analysis: Imperative vs. Declarative Constructs (AI vs. Human)}
\label{tab:gcu-shift}
\begin{tabular}{lccc}
\toprule
\textbf{Category} & \textbf{AI Shift (\%)} & \textbf{Human Shift (\%)} & \textbf{Relative Shift} \\
\midrule
\multicolumn{4}{@{}l}{\textbf{Imperative Signals}} \\
\midrule
On-CPU                     & +0.88\% & -2.57\% & \textbf{+3.46\%} \\
Constructors               & +1.31\% & -0.05\% & \textbf{+1.36\%} \\
Vectors                    & +0.07\% & -0.03\% & \textbf{+0.10\%} \\
Map Lookups                & +0.07\% & -0.01\% & \textbf{+0.07\%} \\
\midrule
\multicolumn{4}{@{}l}{\textbf{Declarative Signals}} \\
\midrule
Libraries (absl/gtl)        & -2.34\% & -1.27\% & \textbf{-1.08\%} \\
\bottomrule
\end{tabular}
\end{table}

\noindent\textbf{Descriptive downstream contrasts.} Figure~\ref{fig:category-outcomes} summarizes the unadjusted downstream contrasts used to motivate the staged models. For review friction, AI-generated changes require significantly more human effort: they receive 1.92$\times$ as many Blocking Threads (reviewer feedback that must be resolved before submission) and 1.39$\times$ as many total Comments. We also observe elevated friction across secondary review metrics, including Reviewer Iterations (1.24$\times$), Time to Merge (1.19$\times$), and Submit Attempts (1.07$\times$). For reliability and process checks, both Sanitizer findings and Build Failure Rates show higher incidences in AI-generated code (medians $\sim$1.3$\times$), while the Revert Rate remains below parity (median $\sim$0.9$\times$).

\textbf{Summary of RQ3.}
The reliability data paints a clear picture: the burden of AI-generated code is not found in acute breakages or reverts. Its comparable build failure rate and superior revert rate suggest that existing automated checks and human reviewers successfully catch fatal errors before, or immediately during, deployment. The true cost of AI generation, therefore, shifts from acute reliability failures to chronic maintenance and operational burdens, manifesting as the interface coupling and compute overheads identified in our static and source-level analysis (RQ2).

\subsection{RQ4: Intervention}\label{sec:intervention}

\begin{figure}
    \centering
    \includegraphics[width=\linewidth]{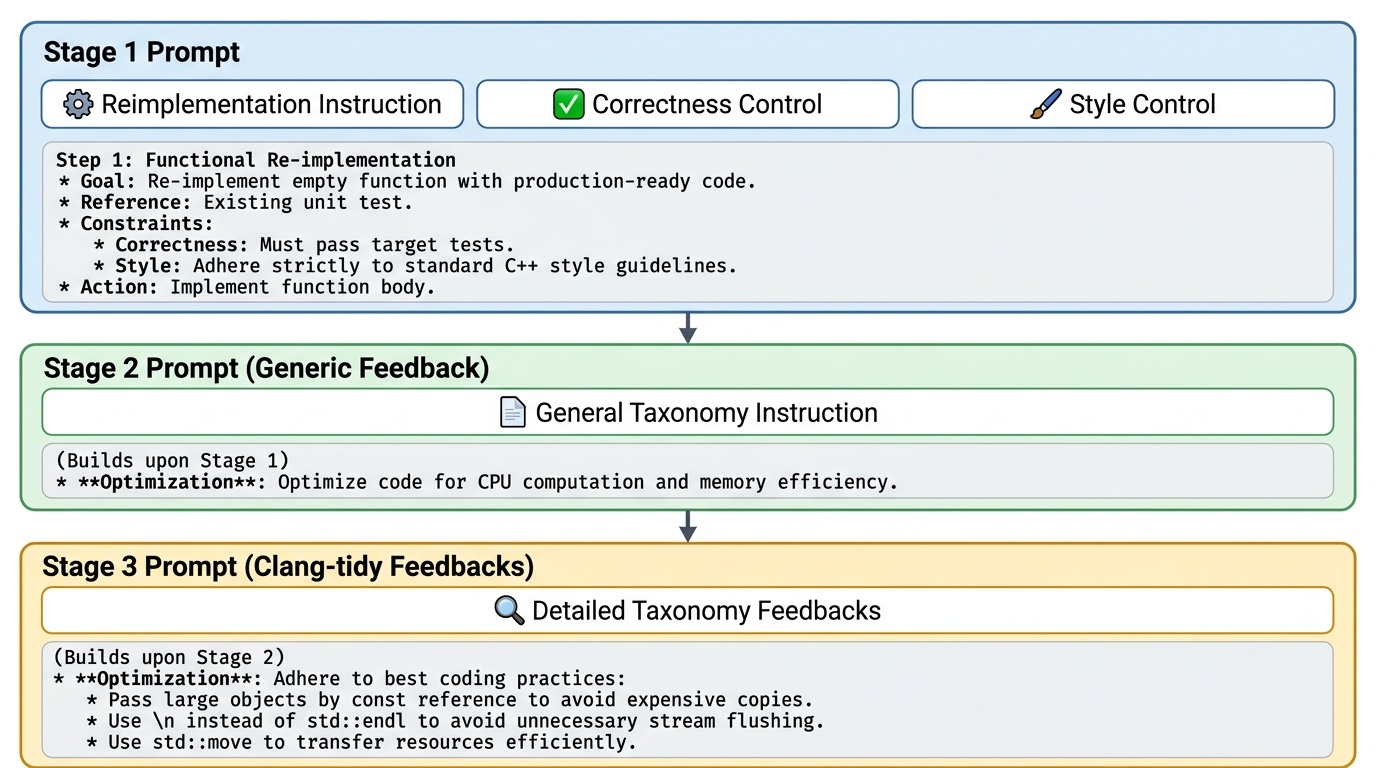}
    \caption{Prompt stages for taxonomy-informed feedback.}
    \label{fig:rq4_prompt}
\end{figure}

RQ4 uses the RQ2 profile as an intervention target. We evaluate taxonomy-informed feedback on the 50-function benchmark described in Section~\ref{subsec:metrics}, where each selected function contains at least one target-category finding from the upstream efficiency profile. Each function is regenerated under 3 prompt stages, as illustrated in Figure~\ref{fig:rq4_prompt}. Stage 1 uses the base reimplementation prompt, Stage 2 adds general feedback, and Stage 3 adds category-specific feedback for the targeted efficiency findings. The benchmark keeps the same function context, reimplementation budget, and validation harness across stages, so changes in findings and $R_{\mathrm{eff}}$ reflect the feedback condition under a fixed task setup.

\textbf{Quantitative impact of taxonomy-informed refinement.} To evaluate whether the concentrated upstream profile can guide feedback, we measured the normalized computational efficiency score ($R_{\mathrm{eff}}$) and the volume of static efficiency findings across the 3 prompt stages. Table~\ref{tab:efficiency_results} reports the results. The baseline in the table stands for the original code snippet before the synthesis. Under the standard drafting instructions (Stage 1), the LLM matched the human baseline finding rate (1.26 versus the 1.28 baseline) with a modest $R_{\mathrm{eff}}$ score of 0.294. 

Stage 3, which adds category-specific feedback, increased the average $R_{\mathrm{eff}}$ score to 0.385 while maintaining low generation variance ($\pm$ 0.043) across three independent runs, a \textbf{31\%} improvement over the prompt without feedback. Stage 3 also reduced the average volume of static efficiency findings to 1.12, an \textbf{11.1\% reduction} relative to Stage 1 and a \textbf{12.5\% reduction} relative to the baseline finding rate. The same RQ2 categories used to characterize submitted code also identify feedback targets that respond in a controlled generation setting.

\begin{table}[t]
\centering
\caption{Taxonomy-informed feedback results.}
\label{tab:efficiency_results}
\footnotesize
\setlength{\tabcolsep}{5pt}
\renewcommand{\arraystretch}{1.04}
\begin{tabular}{@{}lccc@{}}
\toprule
\textbf{Stage} & \textbf{$R_{\mathrm{eff}}$ ($\uparrow$)} & \textbf{Findings ($\downarrow$)} & \textbf{Reduction ($\uparrow$)} \\
\midrule
\textbf{Baseline} & - & 1.28 & - \\
\textbf{Stage 1} & 0.294 ($\pm$0.041) & 1.26 & 1.56\% \\
\textbf{Stage 2} & 0.342 ($\pm$0.041) & 1.22 & 4.69\% \\
\textbf{Stage 3} & \textbf{0.385 ($\pm$0.043)} & \textbf{1.12} & \textbf{12.50\%} \\
\bottomrule
\end{tabular}
\parbox{0.88\columnwidth}{\centering\scriptsize\textit{Note:} Results are averaged across three independent runs.}
\vspace{-10pt}
\end{table}

\noindent\textbf{RQ4 summary.} Taxonomy-informed feedback derived from the upstream profile reduces targeted static efficiency findings on  functions while improving the benchmark efficiency score, providing an initial demonstration that the profile can guide category-specific mitigation.

\section{Discussion}\label{sec:discussion}

We interpret the empirical results as an initial baseline for code quality during the early stages of large-scale AI adoption. The scale result places the quality analysis in an active production workflow, while the C++ results show that the profile remains patterned after developer filtering and review-gated submission. Interface and Coupling Burden and Copy and Allocation Overhead represent much of the positive static gap, which provides reviewers and tool builders with a practical starting point for category-level attention.

% The taxonomy helps interpret raw static checks as developer-facing mechanisms. In RQ2, \texttt{misc-include-cleaner} and \texttt{misc-definitions-in-headers} explain nearly all of Interface and Coupling Burden, while \texttt{runtime-missing-move} explains a large share of Copy and Allocation Overhead.
The taxonomy helps interpret raw static checks as developer-facing mechanisms. In RQ2, \texttt{misc\allowbreak-include\allowbreak-cleaner} and \texttt{misc-definitions\allowbreak-in\allowbreak-headers} explain nearly all of Interface and Coupling Burden, while \texttt{runtime\allowbreak-missing\allowbreak-move} explains a large share of Copy and Allocation Overhead. Those mechanisms are concrete enough for review guidance, analyzer prioritization, and feedback design. They also reveal why target selection should consider both rate contrast and composition, since a category can be overrepresented without dominating review burden, and a common category can dominate findings in both AI-generated and human-written code.

The source-level efficiency measures add a second mechanism for interpreting the profile. The higher loop/API contrast, map-access warnings, container-insertion warnings, and missing-move findings point to code that performs more work locally within generated functions and delegates less work to shared optimized C++ implementations. That pattern helps explain why compute-facing analyses should look beyond low-level instruction throughput. IPC and MIPS show no notable degradation, while normalized compute and heap footprint suffer execution and allocation pressure after deployment.

The feedback study tests whether the profile can support intervention as well as measurement. Stage 3 names targeted efficiency categories selected from the upstream analysis, and the observed reductions show that the same category structure can guide feedback in a controlled generation setting. Practically, the result suggests a tooling path in which static analysis identifies recurring category-level risks and feedback steers generated code away from the same local patterns.

Importantly, the intervention results from RQ4 demonstrate that the issues identified in RQ2 are not hard limits of the underlying models' capabilities. The fact that taxonomy-informed prompting successfully reduced targeted static analysis findings by 11.1\% 
and boosted $R_{eff}$ by 31\% shows that these models are highly malleable. We argue that the observed weaknesses are largely artifacts of historical default system configurations and a lack of context-aware prompting in early implementations. For instance, the models' tendency toward an imperative loop bias vanishes when the system design provides adequate architectural guidance and targeted feedback.

For AI coding tools used in production, pass rate, completion acceptance, and benchmark runtime are insufficient on their own. Evaluation should measure whether generated code is edited away before submission, which static categories remain visible to reviewers, which source-level patterns appear in deployed functions, and whether the observed categories respond to feedback. That evidence standard is consistent with prior work on large-scale analysis ecosystems and practical prioritization of quality signals \cite{sadowski2015tricorder,ivankovic2024productive,frommgen2024reviewcomments,vijayvergiya2024autocommenter}.

\section{Threats to Validity}\label{sec:threats}

\noindent\textbf{Construct validity.} The paper measures AI-generated code through authoring-time provenance observed during authoring and submission. Authoring-time provenance provides stronger measurement than post hoc source-origin attribution, but it does not eliminate all ambiguity \cite{suh2025detecting,wu2025clones,alkaswan2024memorisation,salerno2025remember}. Overlapping AI and human-authoring features on the same bytes, revision mapping from changed bytes to the final submitted snapshot, and tool-specific static checks all introduce measurement choices. The static taxonomy is also intentionally local: it captures line-level issue categories, not every higher-level design or algorithmic property that might later affect compute cost. We address these risks through explicit overlap policies, overlap-excluded sensitivity analyses, frozen check-to-category mappings, support thresholds, and reporting of the unmapped bucket.

\noindent\textbf{Internal validity.} The review and compute analyses are observational. AI-generation share can correlate with task difficulty, repository context, author experience, or review norms. The paper uses sequentially adjusted models that first add code-change measures and then category variables, with controls for time, size, and organizational slice. These steps reduce confounding risk, but they do not convert the study into a causal estimate of AI on quality or cost. The intervention analysis offers a stronger mitigation test, yet it is still narrower than an organization-wide randomized rollout.

\noindent\textbf{Model Mix and Prompt Evolution.} Because individual model  identifiers were masked or aggregated at the data-collection tier,  this paper cannot disentangle whether the observed issues are evenly distributed across all models or heavily driven by earlier, less capable model versions. Furthermore, developer prompting proficiency naturally grew over the course of the 1-year timeline. As a result, older commits may also reflect lower-quality human steering.

\noindent\textbf{Human-in-the-loop confounders and review dynamics:} We also investigated whether standard review metrics—such as prolonged review time or higher iteration counts—correlated with the survival of inefficient AI-generated C++ code, hypothesizing that reviewer fatigue or over-trust might play a measurable role. However, our analysis yielded no clear correlation. The difficulty of isolating these specific architectural trade-offs during standard review suggests that conventional proxy metrics for human effort do not adequately capture the cognitive friction of evaluating AI-generated code. This null result highlights a critical reality in modern workflows: because human reviewers struggle to consistently intercept these localized inefficiencies regardless of review depth, upstream automated interventions—such as the taxonomy-informed prompt feedback evaluated in RQ4—are essential to mitigating post-deployment compute costs.

\noindent\textbf{External validity.} The main static-issue, review, and compute analyses focus on C++ in one large industrial monorepo. That focus is deliberate because the language has strong performance constraints and mature static analysis, but it limits direct transfer to other languages and organizations. The deployed AI authoring tools, rollout timing, and relative mix of interaction modes are also organization-specific, so the observed organization-wide distribution and heterogeneity patterns may not transfer directly to environments with another tool ecology. The results are most directly transferable to organizations with similar review-gated monorepo workflows, authoring-time provenance, mature static analysis, and at least some function-level compute-cost observability \cite{sadowski2015tricorder,sadowski2018modern}. The organization-wide distribution analysis provides broader context across a multi-language production environment, but the category definitions and their associations with downstream outcomes should be interpreted as C++-specific until replicated elsewhere. The taxonomy is a production-grounded characterization of one high-value language, not a universal ontology for all AI-generated code.

\noindent\textbf{Reproducibility and reporting validity.} Some underlying data sources are enterprise-internal, which constrains full release. Company, repository, and tool names are anonymized for double-anonymous review, but the paper retains the contextual details needed to interpret the setting: monorepo-based development, centralized review, multi-language scope, provenance start date, and approximate C++ scale. We further mitigate reporting risk by defining the analytical units, code-change measures, overlap policies, mapping rules, denominators, and sensitivity analyses explicitly, while keeping the taxonomy independent of the internal tool names used to instantiate it. These reporting choices do not fully eliminate replication barriers, but they improve transparency and make partial replication feasible in other industrial settings.

\noindent\textbf{Statistical conclusion validity.} Some issue categories will have low support, and some downstream outcomes may be noisy or highly skewed. The paper addresses this through support thresholds for issue-type reporting, exposure-normalized finding-rate metrics, confidence intervals on category contrasts, staged models, and robustness checks with alternative denominators and overlap policies. Even with these safeguards, small categories, noisy outcome measures, and high-collinearity settings should be interpreted cautiously.

\section*{Data Availability}

This study uses enterprise-internal provenance, review, repository, and production-monitoring data that cannot be released publicly. The anonymous submission includes a replication package for reviewers with the taxonomy definition, category codebook, analysis specifications, model formulas, and selected aggregated statistics needed to support the paper's main claims. The package excludes raw data, person-level records, repository identifiers, service names, tool names, and other internal identifiers. The anonymous submission omits company, repository, and tool names while retaining the contextual details required to interpret the study design.

\section{Conclusion and future works}\label{sec:conclusion}

This paper shows that AI-generated C++ in production has a structured quality profile that can be measured across authoring, submitted code, review, deployed execution, and feedback. Using authoring-time provenance across 3.52 million submitted changes, we find that AI-generated code is widespread but unevenly distributed, is filtered substantially before submission, and shows a C++ static issue-rate gap concentrated in Interface and Coupling Burden and Copy and Allocation Overhead. Source-level efficiency measures point to more explicit local work and lower standard-library/API use, with move/copy and container patterns providing concrete targets for review and tooling.

Beyond direct operational costs, mitigating the compute overhead of AI-generated code is a growing sustainability imperative. As the industry scales AI-assisted authoring, adopting these taxonomy-driven feedback loops will be essential to curbing the broader carbon footprint of enterprise software ecosystems.

A critical driver of the observed compute overhead is the propensity of AI models to generate explicit, imperative logic rather than leveraging optimized, shared libraries. Our source-level analysis demonstrates that AI-generated C++ relies roughly twice as much on local loop constructs and utilizes standard library or API calls 30\% to 40\% less often than human-written code. This behavior is likely symptomatic of generalist LLMs lacking the highly specific context of internal enterprise monorepo structures. Future work should explore the integration of a knowledge base for the coding agents that dynamically inject relevant contexts into the prompt, directly addressing this imperative bias.

The intervention study shows that this profile can guide category-specific feedback on functions, reducing targeted static findings while improving the benchmark efficiency score. Production evaluation of AI-generated code should combine provenance, submitted-change structure, developer-facing issue categories, downstream outcomes, and targeted feedback studies. While our direct injection of taxonomy findings into the re-implementation prompt is effective (RQ4), we plan to extend this feedback in two directions. One is to use a recursive prompt optimization process with specific feedback loops, similar to \cite{blyth2025ieee} with modern prompt tuning like \cite{pryzant2023automatic}. Besides, we plan to incorporate these empirical profiles into a Reinforcement Learning from Execution Feedback (RLEF) pipeline. By building larger synthetic benchmark sets with production-level compute measurements and static analysis findings counts as reward signals, we hope to turn passive, prompt-time interventions into continuous, proactive alignment of the AI system with enterprise performance standards.

This study deliberately focuses on C++ due to its rigorous performance constraints and mature static analysis tooling. However, a critical question for future research is whether these upstream code properties generalize to other strictly typed or performance-critical languages. Further investigation is required to determine if the same imperative biases and library-avoidance patterns manifest in languages like Rust, Go, or Java, or if different language ecosystems produce entirely distinct AI-generated issue taxonomies.

\bibliography{main}

\end{document}